\documentclass[12pt]{article}
\usepackage{epic,eepic}
\def\beq{\begin{equation}}
\def\eeq{\end{equation}}
\def\bea{\begin{eqnarray}}
\def\eea{\end{eqnarray}}

\def\ba{\begin{array}}
\def\ea{\end{array}}
\def\bce{\begin{center}}
\def\ece{\end{center}}

\begin{document}
\begin{titlepage}
\rightline{APCTP-97-13, SNUTP-97-093}
\rightline{UM-TG-196}
\def\today{\ifcase\month\or
        January\or February\or March\or April\or May\or June\or
        July\or August\or September\or October\or November\or December\fi,
  \number\year}
\rightline{hep-th/9707027}
\rightline{revised, July 1998}
\vskip 1cm
\centerline{\Large \bf Branes, Geometry and  $N=1$ Duality}
\centerline{\Large \bf with Product Gauge Groups of $SO$ and $Sp$}
\vskip 1cm
\centerline{\sc Changhyun Ahn$^{a,}$\footnote{ chahn@spin.snu.ac.kr},  
Kyungho Oh$^{b,}$\footnote{ oh@arch.umsl.edu}
and Radu Tatar$^{c,}$\footnote{tatar@phyvax.ir.miami.edu}}
\vskip 1cm
\centerline {{\it $^a$ APCTP, 207-43 Cheongryangri-dong, 
Dongdaemun-gu, Seoul 130-012, Korea and}}
\centerline{{\it Dept. of Physics, Seoul National University,
Seoul 151-742, Korea}}
\centerline{{ \it $^b$ Dept. of Mathematics,
 University of Missouri-St. Louis,
 St. Louis, Missouri 63121, USA}}
\centerline{{\it $^c$ Dept. of Physics, University of Miami,
Coral Gables, Florida 33146, USA}}
 \vskip 2cm
\centerline{\sc Abstract}
\vskip 0.2in
We study $N=1$ dualities in four dimensional supersymmetric gauge theories
as the worldvolume theory of D4 branes with one compact direction
in type IIA string
theory. We generalize the previous work for $SO(N_{c1}) \times Sp(N_{c2})$
with the superpotential $W=\mbox{Tr} X^4$ to the case of $W=\mbox{Tr} 
X^{4(k+1)}$ in terms
of brane configuration. We conjecture that the {\it new} 
dualities for the product
gauge groups of $ SO(N_{c1}) \times Sp(N_{c2}) \times SO(N_{c3})$,
$ SO(N_{c1}) \times Sp(N_{c2}) \times SO(N_{c3}) \times 
Sp(N_{c4}) $ and higher multiple product gauge groups can be obtained by 
reversing the ordering of NS5 branes and D6 branes while 
preserving the linking numbers. We also describe 
the above dualities in terms of wrapping D6 branes around 
3 cycles of Calabi-Yau threefolds in type
IIA string theory.  
The theory with adjoint matter can be regarded as taking multiple copies
of NS5 brane in the configuration of brane or geometric approaches.
\vskip 0.8in
\end{titlepage}
\newpage
\section{Introduction}
In the recent years it has become obvious that Dirichlet(D) 
branes provide an important
tool for studying gauge theories in various dimensions.
There are several approaches to this subject. One is to consider the
compactification of F-theory on elliptic Calabi-Yau fourfolds from 
12 dimensions which gives $N=1$ pure supersymmetric $SU(N_c)$ gauge theories in
four dimensions \cite{KV1}. In \cite{BJPSV}, by studying a configuration with
D3 branes and D7 branes, it has been shown that the local string model
gives rise to $SU(N_{c})$ Yang-Mills theory with matter in the fundamental
representation.

The second approach was initiated by the crucial work of Hanany and Witten
\cite{HW}. They have studied theories with $N=4$ supersymmetry 
in 3 dimensions, obtaining
a nice realization of the mirror symmetry.  By considering type IIB
string theory, they took a configuration which preserves 1/4 of the
supersymmetry, i.e., $N=2$ theory in 4 dimensions or 
$N=4$ theory in 3 dimensions.
Their configurations consisted of {\it parallel} Neveu-Schwarz(NS)5 
branes with D3 branes suspended between them and
D5 branes placed between them. 
A new aspect of brane dynamics was so called ``Hanany-Witten effect",
leading to the creation of a D3 brane whenever a D5 brane and a
NS5 brane are passing through each other. The original explanation was due to
the conservation of the linking number defined as a total magnetic charge for
the gauge field coupled with the worldvolume of the both types of NS and 
D branes.
Another explanation of this effect was given in \cite{BDG} using anomaly
flow argument for two mutually orthogonal D4 branes passing through 
each other where
a fundamental string is created. By a chain of T-dualities and 
S-dualities this reduces to exactly the original ``Hanany-Witten effect". 
Other explanations along the line of \cite{HW} 
were given in \cite{DFK,BGL,deAl}.

>From Hanany and Witten's work, the passage to $N=1$ theories in 4 dimensions
was taken in the important paper of Elizur, Giveon and Kutasov \cite{kuta}.
We call this EGK approach. 
They considered a configuration which preserves 1/8
of the supersymmetry, i.e., $N=1$ theories in 4 dimensions. They considered
type IIA string theory with {\it perpendicular} NS5 branes with D4 branes
suspended between them and
D6 branes placed between them. 
The D4 branes are finite in one direction and the worldvolume dynamics
describes at long distances 
$N=1$ supersymmetric field theories in 4 dimensions. The
configuration of a D4 brane gives  
the gauge group while the D6 branes which are needed for matter give
the global flavor
group. The D4 brane worldvolume gauge coupling 
is related to the finite distance
between the two NS5 branes. That is, the distance between two NS5 branes
in compact direction determines the gauge coupling. 
Using this configuration, they were able to
describe and check the Seiberg's dualities for $N=1$ supersymmetric
gauge theory with $SU(N_{c})$ gauge group with $N_{f}$ flavors 
in the fundamental
representation \cite{Seiberg}. They changed the positions of the two 
NS5 branes in a smooth way, without breaking the supersymmetry and
connected the Higgs phases of the two theories(electric theory and its 
magnetic dual theory). This method sparked
an intense work in obtaining many other dualities for $N=1$ theories
\cite{eva,bh,BSTY,EGKRS,Tatar,bar,John,John1,HZ}. Especially interesting
are the papers of \cite{eva, EGKRS} where the results of \cite{IS, IP} were 
rederived from brane configuration for the gauge groups
$SO(N_{c})$ and $Sp(N_{c})$. They introduced supplementary orientifold
O4 and O6 planes to construct orthogonal and symplectic gauge groups. 
In this case the NS5 branes have to pass over each other
and some strong coupling phenomena have to be considered. For example,
for the case of $SO(N_c)$ gauge group, they consist of
the appearance of supplementary D4 branes or of some D4 branes which
are forced to be stuck at the orientifold O4 plane in order to obtain a smooth 
transition.

The third approach was initiated in the important work of Ooguri and Vafa
\cite{ov}, using the results of \cite{BSV,CDFV}. We call this OV approach.
Instead of working in flat spacetime, they considered the compactification
of IIA string theory on a double elliptically fibered CY threefold. 
They wrapped D6 branes 
around three cycles of CY threefold filling four dimensional
spacetime. The transition between a theory and its magnetic dual appears when
a change in the moduli space of CY threefold occurs. Their result
was generalized in \cite{ca1, ca2, ca3} to various other models which
reproduce field theory results studied previously in 
\cite{Kutasov,ks,intril,ls, BS, Brodie}.

Recently, it has become evident that many important results can be obtained
from strongly coupled type IIA string theory which is given by M theory.
Initiated by the work of Witten \cite{Witten}, this provides an important
way to describe strongly coupled gauge theories.
Starting with a theory of M theory 5 branes in 11 dimensions, a dimensional 
reduction gives  D4 branes (which is an M theory 5 brane 
wrapped over ${\bf S^1}$) and NS5 branes (which is an M theory 5 brane
on ${\bf R^{10} \times S^1}$) in 10 dimensions.   
To obtain D6 branes, one has to use a multiple Taub-NUT space
whose metric is complete and smooth.
The beta function receives a geometrical interpretation. Results of
$N=2$ gauge theories in 4 dimensions with gauge group 
$SU(N_{c})$ appear naturally
in the context of M theory. 
This approach was extended to other gauge groups
in \cite{L3, BSTY1, M3,FS} and to other dimensions \cite{AH,bk,kol}.
Very recently, using a flow from $N=2$ to $N=1$ supersymmetric theories,
non-perturbative results obtained in field theories were
reproduced by using brane configurations in \cite{HOO, bra}. 
Other important results
are also obtained in \cite{witt1, hov, cg}.

In this paper we extend the results of \cite{Tatar, ca3} to the case
of multiple products of $SO$ and $Sp$ gauge groups both in 
EGK and OV approaches. As in \cite{ca3}, the condition to be satisfied 
in the dual theory
is that the flavor groups are given only by D6 branes wrapped on cycles
between {\it pairs of singularity points}, which 
are points where, in the T-dual picture, the NS5 branes which appear
after the T-duality are
parallel. In brane configuration picture, as in \cite{bh} this requires
semi-infinite D4 branes also in our construction. The duality are also
deduced and checked in field theory.

In section 2, we review the result of \cite{Tatar} and rederive it also
from OV approach and generalize to any value of $k$ and for adjoint matter.
In section 3, we extend our results to $SO(N_{c1})\times Sp(N_{c2})
\times SO(N_{c3})$ case for any value of $k$ which appears in the 
superpotential and for adjoint matter.
In section 4, we consider the case of a 4-tuple product gauge group.
In section 5, we generalize to the case of $n$-tuple product gauge group.
Finally in section 6, we conclude our results and comment on the outlook
in the future direction.  

\section{ Duality for $SO(N_{c1}) \times Sp(N_{c2})$ }

The brane configuration we study contains three kinds of
branes in type IIA string theory: NS5 brane
with worldvolume $(x^0, x^1, x^2, x^3, x^4, x^5)$, 
D6 brane with worldvolume $( x^0, x^1, x^2, x^3, x^7, x^8, x^9)$
and D4 brane with worldvolume 
$( x^0, x^1, x^2, x^3, x^6 )$ where
the $x^6$ direction is a finite interval.
We will consider the case of an O4 orientifold in the EGK approach
which is parallel 
to the D4 brane in order to preserve supersymmetry 
and is not of finite extent in $x^6$ direction. The 
D4 brane is the
only brane which is not intersected by O4 orientifold. 
The orientifold gives a spacetime reflection as
$(x^{4},x^{5},x^{7},x^{8},x^{9})\rightarrow (-x^{4},-x^{5},-x^{7},-x^{8},
-x^{9})$
which are reflections on noncompact directions, in addition to the gauging of
worldsheet parity $\Omega$.

On the directions 
where the orientifold is a point, any object which is extended
along them will have a mirror. That is, the NS5 branes have a
mirror in $(x^{4}, x^{5})$ directions
and D6 branes  have one in $(x^{7}, x^{8}, x^{9})$ directions. 
These objects and their mirrors enter only once. 
It would be overcounting to treat
an object and its reflection as separate physical objects.
Therefore we should have the factors of one half in the counting
of the number of physical objects later.
Another important aspect of the orientifold is its charge, given by the charge
of $H^{(6)}=d A^{(5)}$ coming from Ramond Ramond(RR) sector, which is
related to the sign of $\Omega^2$. This is because the charge of the 
orientifold is directly connected to the charge of D branes. 
In the natural normalization,
where the D4 brane carries one unit of this charge, 
the charge of the O4 plane is 
$\pm 1$, for $\Omega^{2}=\mp 1$ in the D4 brane sector.
In this section we will try to understand the field theory results
for $N=1$ dualities \cite{ils} in the context of EGK approach 
first and OV approach later on.

\subsection{ The superpotential W = Tr $ X^{4}$ case}

Let us consider the electric theory for the simplest case
in the sense that each of the three kinds of branes has its single copy. 
The mechanism for duality in the EGK approach is similar that of \cite{bh} except that we
have to consider the orientifold reflection and charges. 
In OV approach, the difference between this section  and  \cite{ov} is 
that we have to twist the doubly elliptic fibration due to
the presence of the product gauge groups with flavors.
We deal with  
$SO(N_{c1})\times Sp(N_{c2})$ gauge group
with $2N_{f1}(2N_{f2})$ flavors in the vector(fundamental) 
representation of $SO(N_{c1})(Sp(N_{c2}))$  
group\footnote{Note that we denote $2N_{f1}$ and $2N_{f2}$ by flavors in
each representation rather than $N_{f1}$ and $N_{f2}$ which were used
in the paper of \cite{ils} in order to avoid the fractional expression when 
we consider its mirrors. We use for simplicity that each factor appearing
in the flavors, representations and gauge groups, corresponds to its own
order.}. 

In the EGK configuration of \cite{Tatar}, this
corresponds to three NS5 branes where we need NS5 branes at arbitrary 
angles in $(x^{4}, x^{5}, x^{8}, x^{9})$ directions so that any two of the
5 branes are not parallel as discussed in 
\cite{bar}.  
We label them by A 5, B 5 and C 5 from left to right on the compact 
$x^{6}$ direction.
There exist $N_{c1}$ D4 branes stretched between A 5 brane and 
B 5 brane which is connected to C 5 brane by $N_{c2}$
D4 branes. 
As discussed in \cite{eva}, the sign of 
the $A^{(5)}$ charge of the orientifold (which is related to the sign of 
$\Omega^2$) flips as one passes a NS5 brane
and flips back again as one passes other NS5 brane. 
If the sign of
$A^{(5)}$ is chosen to be negative between A 5 brane and B 5 brane, 
it will be positive between B 5 brane and C 5 brane. 
For this reason the product gauge group becomes $SO(N_{c1}) \times 
Sp(N_{c2})$(or $Sp(N_{c1}) \times SO(N_{c2})$ 
if we invert the overall sign of the projection). 
We will use these observations in our procedure.
In other words, 
the choice of the gauge group $SO$ or $Sp$ results from the sign of
$\Omega^2$ on the open string sectors. Between A(B) 5 brane and B(C) 5 brane 
we have $2N_{f1}(2N_{f2})$ D6 branes intersecting the $N_{c1}(N_{c2})$
D4 branes. 
Strings stretching between the $2N_{f1}(2N_{f2})$ D6-branes and the 
$N_{c1}(N_{c2})$ D4 branes are 
the chiral multiplets in the vector(fundamental) representation
of $SO(N_{c1})(Sp(N_{c2}))$. 
Strings can also stretch between $N_{c1}$ D4 branes and $N_{c2}$ D4 branes.
We orient the $2N_{f1}(2N_{f2})$ D6 branes in the direction parallel 
to the A(C) 5 brane so there exist chiral multiplets which correspond 
to the motion of D4 branes in between the NS5 and D6 branes, 
as discussed in \cite{bh}. These states are precisely the chiral mesons of
the magnetic dual theory which will be obtained later.

For 
$SO(N_{c1})$ group, we will consider only $N_{c1}/2$ D4 branes, the other
half being their mirrors. For $Sp(N_{c2})$ group, because of the antisymmetric
O4 projections, we have to consider a total even number of D4 branes, i.e.,
$N_{c2}$ D4 branes and their orientifold mirrors. 
When strings are
stretched between the $N_{c1}$ and $N_{c2}$ D4 branes, a field X in the
($N_{c1},2N_{c2}$) representation of product gauge group
$SO(N_{c1}) \times Sp(N_{c2})$ is obtained 
and the superpotential W=Tr$X^{4}$ appears in order to truncate the
chiral ring. 
A nice way to view its appearance 
is to look at our theory from the point of view of
the original $N=2$ supersymmetric theory.
As discussed in \cite{bh,bar}, when all  NS5 branes are parallel we 
have an
$N=2$ SUSY theory while as we rotate the NS5 branes, 
$N=2$ theory is broken and part of the
$N=2$ superpotential appears as a superpotential 
in $N=1$ supersymmetric theory.
Therefore the electric theory is just the one given in
field theory approach \cite{ils}.

{\it The original(electric) picture is the following from left to right:  
Between A 5 brane and 
B 5 brane we have 
$N_{c1}/2$ D4 branes intersecting 
$N_{f1}$ D6 branes
and between B 5 brane and C 5 brane 
we have $N_{c2}$
D4 branes intersecting $N_{f2}$ D6 branes(plus their mirrors).}

In order to find the dual theory,
like in \cite{bh}, first move all the 
$N_{f1}$  D6 branes to the right of all NS5 branes. 
They are intersecting both B 5 brane with C 5 brane. 
Using the ``linking number'' conservation argument (the linking number
for a particular brane must be conserved),
it turns out that each D6 brane has two D4 branes on its left after   
transition.
Similarly by pushing all $N_{f2}$ D6 branes to the left, 
they are crossing both B 5 brane and A 5 brane and  
each D6 brane has two D4 branes on its right after transition.

Because the NS5 branes are trapped at the spacetime orbifold fixed 
point, they cannot avoid their intersection so they should 
meet and there exists a strong coupling singularity when they are
intersecting. When each of B 5 brane and C 5 brane actually meets A 5 brane, 
such a singularity appears. From the field theory point of
view, such a non smooth behavior was expected because for $Sp(N_{c})$ and
$SO(N_{c})$ gauge groups there is a phase transition unlike for $SU(N_c)$ 
gauge group.
In \cite{eva}, the effect of such a singularity was proved to be the appearance
or disappearance of two D4 branes. In \cite{EGKRS} the transition was shown 
to be smooth only when the linking number of both sides of any NS5 brane is 
the same.
For $SO$ gauge group, the procedure is to {\it put} two of the $N_{c1}$ D4
branes  on top of the orientifold plane and {\it break} the other 
$(N_{c1}-2)$ D4 branes on D6 branes , entering in a Higgs phase.
There is no problem to vary smoothly the separation between the two
NS5 branes in $x^6$ direction. 
On the other hand, 
for $Sp$ gauge group a pair of D4 branes and  anti D4 brane plus their mirrors 
are created and anti D4 branes 
cancel the charge difference along the orientifold. 

Let us go on to the magnetic dual theory.
First we move C 5 brane  to the left 
of B 5 brane. In the language of \cite{eva}, two D4 branes must
disappear because we have $Sp$ gauge group between C 5 brane and B 5 brane. 
Then we move C 5 brane to the left of A 5 brane. 
When C 5 brane passes A 5 brane, two D4 branes appear between A 5 brane 
and C 5 brane because we have an $SO$ gauge group,
so totally we have a deficit of two D4 branes between A 5 brane and B 5 brane 
and no extra D4 branes
between C 5 brane and A 5 brane. Now push B 5 brane to the right of A 5 brane. 
When B 5 brane passes A 5 brane (because of the $Sp$ group) there is an extra
 deficit of two D4 branes, but the initial supplementary D4 branes are
changing the orientation and so we have no extra D4 branes in the theory.
In the language of \cite{EGKRS}, for a smooth transition between B 5 brane 
and C 5 brane, we need to create two pairs of D4 branes and anti D4 branes, the
anti D4 branes neutralizing the charge difference along the orientifold as
before. 
The two D4 branes to be put on top of the orientifold, when
C 5 brane passes A 5 brane and B 5 brane passes A 5 brane, annihilate 
the anti D4 branes. The two D4 branes 
which were on top of the orientifold, came from the $N_{c1}$ D4 branes 
connecting A 5 brane and B 5 brane therefore leaving
$(N_{c1}-2)$ D4 branes between A 5 brane and B 5 brane.
Therefore the two D4 branes, which remain after their anti D4 branes vanish,
are added to the $(N_{c1}-2)$ D4 branes. So by smoothing the
transition, we did not produce any D4 or anti D4 branes, as expected from the 
field theory calculation.

{ \it The final(magnetic) picture is the following, from right to left:  
$N_{f1}$ D6 branes are
connected by $2N_{f1}$ D4 branes with A 5 brane. Between A 5 brane and 
B 5 brane we have 
$\widetilde{N}_{c2}$ D4 branes, between B 5 brane and C 5 brane 
we have $\widetilde N_{c1}/2$
D4 branes and to the left of C 5 brane 
we have $N_{f2}$ D6 branes connected by
$2N_{f2}$ branes with C 5 brane(plus their mirrors).}

We use the linking number of A 5 brane to calculate the magnetic
$Sp$ gauge group, $\widetilde{N}_{c2}$. 
In the original electric picture it was equal to $- N_{f1}/2- N_{f2}/2+
N_{c1}/2$ where we followed 
conventions: NS5 branes or D6 branes to the left(right) of the brane
we are considering give a contribution of $1/2(-1/2)$ to the linking number 
while D4 branes to the left(right) give a contribution of $-1(1)$.
In the 
magnetic picture the linking number becomes
$-N_{f1}/2+ N_{f2}/2-\widetilde{N}_{c2}+2N_{f1}$. By making them equal, 
we obtain $\widetilde{N}_{c2}=2N_{f1}+N_{f2}-N_{c1}/2$.
For the B 5 brane, the conservation of the linking number gives the relation 
$-N_{f2}/2+N_{f1}/2+N_{c2}-N_{c1}/2=\widetilde{N}_{c2}-
\widetilde{N}_{c1}/2-N_{f1}/2+
N_{f2}/2$. Then we obtain $\widetilde{N}_{c1}=2N_{f1}+4N_{f2}-2N_{c2}$. 
The values for
$\widetilde{N}_{c1}$ and $\widetilde{N}_{c2}$ coincide with precisely 
the ones obtained in \cite{ils} by remembering that we started with
even number of flavors $2N_{f1}$ and $2N_{f2}$.
The magnetic theory obtained by inverting the order of
NS5 branes can be described as a theory with the gauge group 
$SO(\widetilde{N}_{c1})\times
Sp(\widetilde{N}_{c2})$ with $\widetilde{N}_{c1}=2N_{f1}+4N_{f2}-2N_{c2}$ and
$\widetilde{N}_{c2}=2N_{f1}+N_{f2}-N_{c1}/2$. 
>From the brane configuration discussed above, the field contents of the theory
are as follows: 
the gauge group $SO(\widetilde{N}_{c1})\times Sp(\widetilde{N}_{c2})$
with $2N_{f2}(2N_{f1})$ fields in the vector(fundamental) 
representation of $SO(\widetilde{N}_{c1})(Sp(\widetilde{N}_{c2}))$
and a field Y in the ($\widetilde{N}_{c1},2\widetilde{N}_{c2}$) 
representation of the product gauge group where the chiral 
mesons of the dual theory which have the same form as in \cite{ils}.
 
Now we want to describe the above duality in the context of OV
approach and to see what is the correspondence between them. 
In \cite{ov}, they obtained an $N=1$ theory in $d=4$ living on the (3+1)-part
of the worldvolume of D6 branes by 
partially wrapping the D6 branes around vanishing 
real 3-cycles in  a local model of a doubly elliptic fibered
Calabi-Yau threefold in type IIA string theory. 
 
In order to study  $SO(N_{c1})\times Sp(N_{c2})$ gauge theories with flavors,
 we consider 
a twisted  local model of a doubly elliptic fibered CY threefold given by
\bea
\label{cy}
(x&+&[m_1 +m_2(\mu_1 z-\mu_2)(z- a_1)(z-a_2)] x')^2+\nonumber\\
(y&+& [m_1 +m_2(\mu_1 z-\mu_2)(z- a_1)(z-a_2)] y')^2 
= -(z-a_1)(z-a_2)(z-c_2)(z-c_1)\nonumber\\
x'^2&+&y'^2=-z\nonumber\\
\mu_1& =& \frac{1}{c_1}\left\{\frac{1}{(c_1-a_1)(c_1 -a_2)}  -  
\frac{1}{2} \left[\frac{1}{(c_1-a_1)(c_1 -a_2)} 
                          -\frac{1}{(c_2-a_1)(c_2 -a_2)}\right]\right\}
\nonumber\\
\mu_2 &=&  \frac{1}{2} \left[\frac{1}{(c_1-a_1)(c_1 -a_2)} 
                          -\frac{1}{(c_2-a_1)(c_2 -a_2)} \right]
\eea
where $a_i, c_i$'s are real numbers with $a_1< a_2 <0 < c_2< c_1$.
Here $m_1$ and $m_2$ are arbitrary real numbers. When they are zero, we get
back the usual local model of a CY threefold. 
Via the projection $(x,y,x',y',z) \to z$, we may regard the above local 
description as an approximation of 
a doubly elliptic fibered CY threefold  over the z-plane near
its degeneration. The general fiber is isomorphic to
${\bf C^{*}\times C^{*}}$, but we cannot split the
fibers uniformly (i.e. independent of $z$)
into a product of two ${\bf C^{*}}$ because 
 the first ${\bf C^{*}}$ is not embedded in $(x, y)$-space even though the second ${\bf C^{*}}$
 is still embedded in
$(x', y')$-space.
The CY threefold is smooth, but the fiber
over the z-plane 
acquires $A_1$ singularities as $z$ approaches $a_i, c_i$ or 0. 
For a fixed $z$ away from $a_i, c_i$ or $0$, there exist nontrivial 
$S^1$'s in each of ${\bf C^{*}}$ which vanishes as $z$ approaches 
to $a_i, c_i$ or 0 so that the union of these vanishing $S^1$'s 
over a real line segment near
$a_i, c_i$ or 0 will look like a `thimble'.
Thus the cycles over the real line segments $[a_2, a_1], [c_2, c_1]$ are
$S^3$'s and the cycles over the real line segment $[a_1, 0], [0, c_2]$
are $S^2 \times S^1$'s.

Now we orientifold this configuration by combining the complex conjugation
\bea
(x, y, x',y', z) \rightarrow (x^{*}, y^{*}, x'^{*}, y'^{*}, z^{*})
\eea
with exchange of left- and right-movers on the worldsheet.
Thus the orientifold 6-space (i.e. the fixed point set)
 is the real CY threefold defined by the
equation (\ref{cy}) times the uncompactified $(x_0, x_1, x_2, x_3)$ space.
As it is shown in \cite{ov}, one can see that the degeneration of
the fiber at each $a_i, c_i$ or 0 is replaced by one NS 5 brane
after performing the T-duality on each ${\bf C^{*}}$ of
the double elliptic fibration. 
We choose coordinates so that NS 5 branes arising at $a_i$ or $c_i$'s when
$m_i =0$ in the equation (\ref{cy})
 are parallel to the $x^0, \ldots , x^3, x^4, x^5$ plane,
and NS 5 brane arising at 0 is parallel to the 
$x^0, \ldots , x^3, x^8, x^9$ plane. But for generic $m_i$, NS 5 branes
at $a_i$'s  are parallel but not parallel to NS 5 brane at 0 or $c_i$'s.
Also NS 5 branes at $c_i$'s are parallel, 
but not parallel to NS 5 brane at 0 or $a_i$'s..
Being common transverse direction to all NS 5 branes,
$(x^{6}, x^{7})$ will be regarded as  real and imaginary
parts of $z$ coordinate. 

The contribution of the O6 plane to the D6 brane charge is determined by
noting that O9 plane carries  $(-16)$ units of D9 charge (See, for example,
\cite{polchin}). Then we compactify
on a three torus $T^3$ and T-dualize. We obtain  eight O6 planes with
the same total charge as before, that is, $(-16)$ units of D6 brane charge.
Therefore, each O6 plane carries $(-2)$ units of D6 brane charge.
Because the contribution of the O6 plane to D6 brane charge
comes from a diagram with a single crosscap, going from $SO$ to $Sp$
involves the sign change for diagrams with odd number of crosscaps.
This means O6 plane will carry $+2$ units of D6 brane charge for the $Sp$
case.

{ \it We consider $(N_{c1}/2-2)$ D6 brane charge
on the cycle $[a_{1}, a_2]$, $(N_{c2}+2)$ D6 brane charge 
on the cycle $[c_2, c_{1}]$,
$N_{f1}$ D6 brane charge on the cycle $[a_2, 0]$ and 
$N_{f2}$ D6 brane charge on
the cycle $[0, c_1]$ as in the following Figure 1.}

\begin{figure}[htbp]

\setlength{\unitlength}{0.00068in}
\begin{picture}(7473,1995)(-500,-10)
\thicklines
\put(912.000,561.000){\arc{1950.000}{3.5364}{5.8884}}
\put(2712.000,1311.000){\arc{1950.000}{0.3948}{2.7468}}
\put(4512.000,1311.000){\arc{1950.000}{0.3948}{2.7468}}
\put(6312.000,561.000){\arc{1950.000}{3.5364}{5.8884}}
\path(12,936)(7212,936)
\put(12,636){$\mathbf{a_1}$}
\put(1812,636){$\mathbf{a_2}$}
\put(3612,636){$\mathbf{0}$}
\put(5412,636){$\mathbf{c_2}$}
\put(7212,636){$\mathbf{c_1}$}
\put(612,1836){$\mathbf{\frac{ N_{c_1}}{2}-2}$}
\put(6012,1836){$\mathbf{ N_{c_2}+2}$}
\put(2412,36){$\mathbf{ N_{f_1}}$}
\put(4212,36){$\mathbf{ N_{f_2}}$}
\end{picture}
\caption{}
\label{fig1}
\end{figure}

 
To make the flavor groups global we need to push $a_{2}$ and 
$c_{2}$ to infinity.
To do that we have to push $a_2$ to the right (resp.  $c_2$ to the left)
since NS branes at $a_i$'s (resp. $c_i$'s) are parallel in the T-dual picture.
The main geometric difficulty here is that we cannot lift the degeneration
points off the real axis because it is frozen by the orientifolding.
As two of these degeneration points collide, it will create a singular
CY threefold. Physically, this means that we have to go through a strong 
coupling region. Rather than directly investigating the situation, we
perform T duality and appeal to the linking number argument as before.

We now have five NS5 branes placed at each of the previous five
points. The ones at $a_{1}$ and $a_{2}$ (resp. $c_1$ and $c_2$) are 
parallel and are not parallel to any others. 
The NS5 branes
at $a_{2}$ and $c_{2}$ play the role of the $N_{f1}$ and $N_{f2}$ D6 branes
respectively
in the brane configuration picture we have seen where by
moving $N_{f1}$ D6 brane from right to left with respect to a NS5 brane, 
$N_{f1}$ D4 brane 
will appear to the left of the $N_{f1}$ D6 brane. 
Our claim here is that when we move
the NS5 branes sitting at $c_{2}$ from right to left with respect to the one
sitting at $0$ or $a_2$, the same amount of D4 branes will appear. But 
no more $D4$ branes will be created when $c_2$ passes thru from $a_2$ to $a_1$
because the cycle type has changed once $c_2$ passes through $a_2$ and
the NS branes at $a_1$ and $a_2$ are parallel.
The creation of $D4$ branes 
 has been explained in \cite{ov} geometrically. We will give another possible
explanation in the appendix.

Going back (by T-duality) to the original picture, 
this means that when we move
$c_{2}$ to the left of $a_{i}$ and $0$ then 
$2N_{f2}$ supplementary D6 branes are wrapped
on $[c_{2}, a_{1}]$. The same argument tells us that 
when we move $a_{2}$ to the
right of $c_{1}$, then $2N_{f1}$ supplementary D6 branes are wrapped on
$[c_{1}, a_{2}]$.  
Now we want to move to other point in the moduli of the CY threefolds 
and end up
with a configuration where again the degeneration points are 
along the real axis
in the $z$-plane, but the order is changed to 
$(c_{2}, c_{1}, 0, a_{1}, a_{2})$.
We decide the number of D6 branes wrapping on $[c_1, 0]$ and $[0, c_2]$
by linking number argument.

{ \it Thus we obtain the final configuration of points
ordered as $(c_{2}, c_{1}, 0, a_{1}, a_{2})$ with $2N_{f2}$ D6 branes 
wrapped on
$[c_{2}, c_{1}]$, $(N_{f1} + 2N_{f2}-N_{c2}-2)$ D6 branes wrapped on
$[c_{1}, 0]$, $(2N_{f1} + N_{f2} -N_{c1}/2+2) $ D6 branes wrapped on 
$[0, a_{1}]$ and $2 N_{f1}$ D6 branes wrapped on $[a_{1}, a_{2}]$ as in the
Figure 2.}


\begin{figure}[htbp]

\setlength{\unitlength}{0.0125in}
\begin{picture}(332,110)(-80,-10)
\put(40.000,60.000){\arc{100.000}{0.6435}{2.4981}}
\put(120.000,0.000){\arc{100.000}{3.7851}{5.6397}}
\put(280.000,60.000){\arc{100.000}{0.6435}{2.4981}}
\put(200.000,30.000){\arc{80.000}{3.1416}{6.2832}}
\path(0,30)(320,30)
\put(0,10){$\mathbf{ c_2}$}
\put(80,10){$\mathbf{ c_1}$}
\put(160,10){$\mathbf{ 0}$}
\put(240,10){$\mathbf{ a_1}$}
\put(320,10){$\mathbf{ a_2}$}
\put(25,0){$\mathbf{ 2N_{f_2}}$}
\put(265,0){$\mathbf{ 2N_{f_1}}$}
\put(75,52){$\mathbf{ N_{f_1}+2N_{f_2}-N_{c_2}-2}$}
\put(155,72){$\mathbf{ 2N_{f_1}+N_{f_2}-\frac{N_{c_1}}{2}+2}$}
\end{picture}

\caption{}
\label{fig2}
\end{figure}


This picture describes the magnetic dual theory which has a gauge group
$SO(2N_{f1} + 4N_{f2} - 2N_{c2})\times Sp(2N_{f1} + N_{f2} - N_{c1}/2)$ with
$2N_{f2}( 2N_{f1} )$ flavors in the vector(fundamental) representation of 
the $SO(Sp)$ gauge group.
In order to obtain global flavor groups, we push $a_{2}$ and $c_{2}$
to infinity. We have singlets which correspond to the mesons of the
original electric theory and interact with the dual quarks through the
superpotential in the magnetic theory.

\subsection{ The superpotential W = Tr $ X^{4(k+1)}$ case}

We now move to a more general form of superpotential. In brane configuration,
it is straightforward to show that 
the previous arguments can be generalized by taking multiple 
copies of NS5 branes.

{\it This theory corresponds to (from left to right) $(2k+1)$ 
NS5 branes with the 
orientation of C connected to a single B 5 brane by $N_{c2}$ D4 branes.
The single B 5 brane is connected to $(2k+1)$ NS5 branes with the same 
orientation as A  by
$N_{c1}/2$ D4 branes.
$N_{f1}$ and $N_{f2}$ D6 branes intersect the $N_{c1}$ and $N_{c2}$ 
D4 branes, respectively (plus their mirrors).}

As done in previous subsection, we move all the $N_{f1}$  D6 branes to the 
left of all NS5 branes. 
They intersect both B 5 brane and $(2k+1)$ C 5 branes. 
Using the linking number conservation argument
each D6 brane has  $2(k+1)$ D4 branes on its right after the   
transition. Similarly, by moving all $N_{f2}$ D6 branes to the right, 
they are intersecting both
B 5 brane and $(2k+1)$ A 5 brane and from the conservation of linking number 
each D6 brane has $2(k+1)$ D4 branes on its left after transition.

{\it The final picture is the following, from left to right:  
$N_{f1}$ D6 branes 
connected by $2(k+1) N_{f1}$ D4 branes with $(2k+1)$ A 5 branes. 
Between $(2k+1)$ A 5 branes and 
B 5 brane we have 
$\widetilde{N}_{c2}$ D4 branes, between B 5 brane and $(2k+1)$ C 5 branes 
we have $\widetilde N_{c1}/2$
D4 branes and to the right of $(2k+1)$ C 5 branes 
we have $N_{f2}$ D6 branes, connected by
$2(k+1) N_{f2}$ D4 branes with $(2k+1)$ C 5 branes(plus their mirrors).}

We apply the linking number conservation argument for A 5 brane to calculate 
$\widetilde{N}_{c2}$. In the original electric picture 
it is equal to $-(2k+1) N_{f1}/2-(2k+1) N_{f2}/2+
N_{c1}/2$. In the magnetic picture the linking number is
$-(2k+1) N_{f1}/2+(2k+1) N_{f2}/2-\widetilde{N}_{c2}+2(k+1) N_{f1}$. 
Making them equal, we obtain
$\widetilde{N}_{c2}=2(k+1) N_{f1}+(2k+1) N_{f2}-N_{c1}/2$.
For the B 5 brane, the conservation of the linking number gives 
$-N_{f2}/2+N_{f1}/2+N_{c2}-N_{c1}/2=\widetilde{N}_{c2}-
\widetilde{N}_{c1}/2-N_{f1}/2 + N_{f2}/2$. 
We obtain $\widetilde{N}_{c1}=2(2k+1) N_{f1}+ 4(k+1) N_{f2}-2N_{c2}$. 
The values for
$\widetilde{N}_{c1}$ and $\widetilde{N}_{c2}$ coincide with 
the ones discussed in \cite{ils} exactly.

In the geometrical picture, after a T-duality, this would
mean that instead of having single NS 5-brane at $a_{1}$ we have
$(2k+1)$ NS5 branes and instead of having one NS 5-brane at $c_{1}$ we have
$(2k+1)$ NS5 branes. In order to handle this problem we may regard
$(2k+1)$ NS5 branes at one point as  $(2k+1)$ 
different points where each NS5 brane corresponds to  each point.
This point of view was introduced in \cite{ca1} for the gauge group
$SU(N_{c})$ with adjoint matter.
For simplicity, let us consider first the case $k=1$(Remind that
the theory of the previous subsection corresponds to the case of $k=0$). 
We have to twist the doubly elliptic fibration again so that
in T-dual picture only the branes $a_{11}, a_2$ (resp. $c_2, c_{11}$) 
are parallel and all others are at arbitrary angles.

{\it On the real axis we will
have from left to right: $(a_{13}, a_{12}, a_{11}, a_{2}, 0, 
c_{2}, c_{11}, c_{12}, c_{13})$,
where instead of $a_{1}$ (resp. $c_1$) we have $a_{11}, a_{12}, a_{13}$
(resp. $c_{11}, 
c_{12}, c_{13}$). 
Using the idea of \cite{ca1,ca2}, we have D6 brane charges $(N_{c1}/2-2)$
on $[a_{11}, a_2]$, 
$(N_{c2}+2)$ on $[c_2, c_{11}]$.
 Besides we also have $N_{f1}$ D6 brane charge on $[a_2, 0]$ 
and $N_{f2}$ D6 brane charge on $[0, c_2]$ and a single
D6 brane charge on $[a_{13}, a_{12}], [a_{12}, a_{11}], [c_{11},
c_{12}]$ and $[c_{12}, c_{13}]$ as in Figure 3. }

\begin{figure}[htbp]

\setlength{\unitlength}{0.0008in}
\begin{picture}(7550,1695)(-200,-10)
\put(1812.000,636.000){\arc{1200.000}{3.1416}{6.2832}}
\put(3012.000,1086.000){\arc{1500.000}{0.6435}{2.4981}}
\put(4212.000,1086.000){\arc{1500.000}{0.6435}{2.4981}}
\put(5412.000,636.000){\arc{1200.000}{3.1416}{6.2832}}
\put(312.000,636.000){\arc{600.000}{3.1416}{6.2832}}
\put(912.000,636.000){\arc{600.000}{3.1416}{6.2832}}
\put(6312.000,636.000){\arc{600.000}{3.1416}{6.2832}}
\put(6912.000,636.000){\arc{600.000}{3.1416}{6.2832}}
\path(12,636)(7212,636)
\put(1512,1486){$\mathbf{\frac{N_{c1}}{2} -2}$}
\put(2912,36){$\mathbf{N_{f1}}$}
\put(4112,36){$\mathbf{N_{f2}}$}
\put(5112,1486){$\mathbf{N_{c2} +2}$}
\put(2412,336){$\mathbf{ a_2}$}
\put(3612,336){$\mathbf{ 0}$}
\put(4812,336){$\mathbf{ c_{2}}$}
\put(312,1086){$\mathbf{1}$}
\put(912,1086){$\mathbf{1}$}
\put(6312,1086){$\mathbf{1}$}
\put(6912,1086){$\mathbf{1}$}
\put(12,336)  {$\mathbf{ a_{13}}$}
\put(612,336) {$\mathbf{ a_{12}}$}
\put(1212,336){$\mathbf{ a_{11}}$}
\put(6012,336){$\mathbf{ c_{11}}$}
\put(6612,336){$\mathbf{ c_{12}}$}
\put(7212,336){$\mathbf{ c_{13}}$}
\end{picture}

\caption{}
\label{fig3}
\end{figure}


Now
we move $c_{2}$ to the left  of $a_{11}, a_{12}, a_{13}$ and 
$a_{2}$ to the right of
$c_{11}, c_{12}, c_{13}$.
To apply the observation made in the previous subsection, we
perform first a T-duality
in $(x^{4}, x^{5}, x^{8}, x^{9})$ directions. In the T-dual language,
moving $c_{2}$ to the left of $a_{11}, a_{12}, a_{13}$ means moving the 
NS5 brane sitting at $c_{2}$ to the left with respect to the 
NS5 branes sitting
at $a_{11}, a_{12}$ and $a_{13}$. When $c_{2}$
passes $0, a_2$ and $ a_{1i}( i=2, 3)$, $N_{f2}$ supplementary D6 branes 
are created. 
Going back to the
original geometrical picture, we have
$4N_{f2}$ D6 branes wrapped on $[c_{2}, a_{13}]$. 
The same thing appears when we move $a_{2}$ to the right of $c_{11}, 
c_{12}$ and $c_{13}$. 
Note that the branes between $a_{1i}$'s or $c_{1i}$'s
disappear in the limit $a_{11} =a_{12} = a_{13}$ or $c_{11} =
c_{12}= c_{13}$.
We now want to move 
to another point of moduli space of the CY threefold by changing the positions 
of $a_{11}=a_{12}=a_{13}=a_{1}$ and $c_{11} = c_{12} = c_{13}=
c_{1}$. We push $c_{1}$
and $0$ to the right of $a_{1}$ to obtain the magnetic theory. 
The numbers of D6 branes wrapped on $[c_1, 0]$ and $[0, a_1]$ are decided
by their linking numbers.

{\it The
final configuration is $(c_{2}, c_{1}, 0, a_{1}, a_{2})$ with 
$4N_{f2}$ D6 branes wrapped on $[c_{2}, c_{1}]$, $(3N_{f1}+4N_{f2}-
N_{c2}-2)$
D6 branes wrapped on $[c_{1}, 0]$, $(4N_{f1} + 3N_{f2}- N_{c1}/2+2)$ 
D6 branes
wrapped on $[0, a_{1}]$ and $4N_{f1}$ D6 branes wrapped on 
$[a_{1}, a_{2}]$ as in Figure 4.}


\begin{figure}[htbp]

\setlength{\unitlength}{0.0125in}
\begin{picture}(332,110)(-80,-10)
\put(40.000,65.000){\arc{100.000}{0.6435}{2.4981}}
\put(120.000,5.000){\arc{100.000}{3.7851}{5.6397}}
\put(280.000,65.000){\arc{100.000}{0.6435}{2.4981}}
\put(200.000,35.000){\arc{80.000}{3.1416}{6.2832}}
\path(0,35)(320,35)
\put(0,15){$\mathbf{ c_2}$}
\put(80,15){$\mathbf{ c_1}$}
\put(160,15){$\mathbf{ 0}$}
\put(240,15){$\mathbf{ a_1}$}
\put(320,15){$\mathbf{ a_2}$}
\put(20,5){$\mathbf{ 4N_{f_2}}$}
\put(70,57){$\mathbf{ 3N_{f_1}+4N_{f_2}-N_{c_2}-2}$}
\put(150,77){$\mathbf{ 4N_{f_1}+3 N_{f_2} - \frac{N_{c_1}}{2} +2}$}
\put(260,5){$\mathbf{ 4N_{f_1}}$}
\end{picture}

\caption{}
\label{fig4}
\end{figure}


This gives a field theory with the gauge group 
$SO(2(3N_{f1}+4N_{f2})-2N_{c2})\times Sp(4N_{f1}+3N_{f2}-N_{c1}/2))$ with 
$2N_{f2}(2N_{f1})$
flavors in the vector(fundamental) of the $SO(\widetilde{N}_{c1})(Sp(
\widetilde{N}_{c2}))$ gauge group where $\widetilde{N}_{c1}$ and $
\widetilde{N}_{c2}$ are the same as those for $k=1$ 
given in brane configuration.

The generalization to arbitrary $k$ becomes now obvious. Instead of 
$a_{1}(c_1)$ we 
take $(2k+1)$ singular points like as $a_{11}, a_{12}, \cdots, a_{1(2k+1)}(
c_{11}, c_{12}, \cdots, c_{1(2k+1)})$. 
The points $a_{2}, 0 $ and $ c_{2}$ will still remain
in a single copy. Now we wrap a single D6 brane  on $[a_{1i}, a_{1i+1}]$ 
for each $i$  and on $[c_{1j}, c_{1j+1}]$ for each $j$. Also
we wrap $N_{f1}$ on $[0, a_{2}]$ and $N_{f2}$ on $[c_{2}, 0]$. When we 
move $a_{11}, \cdots, a_{1(2k+1)}$ to the right of $c_{2}$, 
for every transition,
$N_{f2}$ supplementary D6 branes will appear. 
Eventually we have $2(k+1)N_{f2}$ D 6-branes wrapped on
$[c_{2}, a_{11}]$ and so on. 
The same thing happens when we push $c_{11}, \cdots, c_{1(2k+1)}$
to the left of $a_{2}$ and we end up with $2(k+1)N_{f1}$ branes wrapped on 
$[c_{11}, a_{2}]$. Now we identify $a_{1i}(c_{1j})$ with $a_{1}(c_{1})$
like as $a_{11} = \cdots = a_{1(2k+1)}= a_{1}$ and 
$c_{11}= \cdots = c_{1(2k+1)}= c_{1}$ and 
change again the positions of $0$ and $c_{1}$ from 
the right to the left of $a_{1}$ to obtain the magnetic dual description. 
This has a gauge group 
$SO(2(2k+1)N_{f1} + 4(k+1)N_{f2}- 2N_{c2})\times 
Sp(2(k+1)N_{f1} + (2k+1)N_{f2}-N_{c1}/2)$ which is exactly the same as that 
obtained in brane configuration. 

\subsection{ The addition of adjoint matter }

Now we can go on one further step by taking multiple copies of the middle
NS5 brane.
In brane configuration, we claim that the model with adjoint matter 
is described by  $(2k+1)$ A 5 branes connected to $(2k+1)$ B 5 branes
by $N_{c1}$ D4 branes. The $(2k+1)$ B 5 branes are connected to $(2k+1)$
C 5 branes by $N_{c2}$ D4 branes. The $N_{f1}$ and $N_{f2}$ D6 branes
intersect the $N_{c1}$ and $N_{c2}$ D4 branes respectively.

We use the linking number of A 5 brane to calculate $\widetilde{N}_{c2}$ 
as usual. In the original electric picture 
it was equal to $-(2k+1) N_{f1}/2-(2k+1) N_{f2}/2+
N_{c1}/2$. In the magnetic picture the linking number can be written as
$-(2k+1) N_{f1}/2+(2k+1) N_{f2}/2-\widetilde{N}_{c2}+2(2k+1) N_{f1}$. 
The conservation of linking number allows us to have
$\widetilde{N}_{c2}=2(2k+1) N_{f1}+(2k+1) N_{f2}-N_{c1}/2$.
For the B 5 brane, the conservation of the linking number gives the
relation 
$-(2k+1)N_{f2}/2+(2k+1)N_{f1}/2+N_{c2}-N_{c1}/2=\widetilde{N}_{c2}-
\widetilde{N}_{c1}/2-(2k+1)N_{f1}/2+
(2k+1)N_{f2}/2$. 
>From this we obtain $\widetilde{N}_{c1}=2(2k+1) N_{f1}+ 4(2k+1) N_{f2}-
2N_{c2}$. 

We now go to the geometrical picture and see how the 
geometric configuration appears. Let us take $k=1$ for 
simplest case:
$(a_{13}, a_{12}, a_{11}, a_{2}, b_{1}, 0, b_{2}, c_{2}, c_{11}, c_{12},
c_{13})$.
Note that $a_{2}$ and $ c_{2}$ are always in a single copy. We now wrap
$(N_{c1}/2-2)$ D6 brane charge
on $[a_{11}, a_{2}]$, 
$N_{f1}$ on $[a_{2}, b_{1}]$, 
1 on $[b_{1}, 0]$, 1 on $[0, b_2]$,
$N_{f2}$ on $[b_{2}, c_{2}]$, $(N_{c2}+2)$
on $[c_{2}, c_{11}]$, $1$ on $[c_{11}, c_{12}]$ and 1 on $[c_{12}, c_{13}]$. 
If we take the limit $b_{1}, b_{2} \rightarrow 0$, 
we have now $3 N_{f1}$ D6 branes 
wrapped on $[a_{2}, b_{1} = b_{2}= 0]$ and $3N_{f2}$ D6 branes on 
$[b_{1} = b_{2}=0, c_{2}]$. 
Again the configuration looks very much alike the
previous configurations. We draw the schematic picture in the Figure 5.

\begin{figure}[htbp]

\setlength{\unitlength}{0.00068in}
\begin{picture}(8424,1359)(0,-10)
\put(312.000,711.000){\arc{600.000}{3.1416}{6.2832}}
\put(912.000,711.000){\arc{600.000}{3.1416}{6.2832}}
\put(1812.000,1161.000){\arc{1500.000}{0.6435}{2.4981}}
\put(3012.000,1161.000){\arc{1500.000}{0.6435}{2.4981}}
\put(3912.000,711.000){\arc{600.000}{3.1416}{6.2832}}
\put(4512.000,711.000){\arc{600.000}{3.1416}{6.2832}}
\put(5412.000,1161.000){\arc{1500.000}{0.6435}{2.4981}}
\put(6612.000,1161.000){\arc{1500.000}{0.6435}{2.4981}}
\put(7512.000,711.000){\arc{600.000}{3.1416}{6.2832}}
\put(8112.000,711.000){\arc{600.000}{3.1416}{6.2832}}
\path(12,711)(8412,711)
\put(312,1236){$\mathbf{1}$}
\put(912,1236){$\mathbf{1}$}
\put(1587,111){$\mathbf{\frac{N_{c1}}{2} -2}$}
\put(2862,111){$\mathbf{N_{f1}}$}
\put(3837,1236){$\mathbf{1}$}
\put(4437,1236){$\mathbf{1}$}
\put(5112,111){$\mathbf{N_{f2}}$}
\put(6312,111){$\mathbf{N_{c2}+2}$}
\put(7437,1236){$\mathbf{1}$}
\put(7962,1236){$\mathbf{1}$}
\put(12,411){$\mathbf{ a_{13}}$}
\put(612,411){$\mathbf{ a_{12}}$}
\put(1212,411){$\mathbf{ a_{11}}$}
\put(2412,411){$\mathbf{ a_2}$}
\put(3612,411){$\mathbf{ b_{1}}$}
\put(4212,411){$\mathbf{ 0}$}
\put(4812,411){$\mathbf{ b_{2}}$}
\put(6012,411){$\mathbf{ c_{2}}$}
\put(7212,411){$\mathbf{ c_{11}}$}
\put(7812,411){$\mathbf{ c_{12}}$}
\put(8412,411){$\mathbf{ c_{13}}$}

\end{picture}

\caption{}
\label{fig5}
\end{figure}


We now move $c_{2}$ to the left of $a_{11}, a_{12}$ and  $a_{13}$
and $a_{2}$
to the right of 
$c_{11}, c_{12}$ and $c_{13}$. 
In this case, we apply again the observation 
about the creation of supplementary D6 branes we have discussed before. 
We obtain $6 N_{f2}$ D6 branes wrapped
on $[c_{2}, a_{13}]$ and
$6 N_{f1}$ wrapped on $[c_{13}, a_{2}]$. 

We now move to other point in the moduli of CY threefolds and end up with
the magnetic dual theory. 
The
procedure is the same as before and we obtain the magnetic
theory with the gauge group $SO(6N_{f1} + 12N_{f2} - 2N_{c2})\times
Sp(6N_{f1} + 3N_{f2} - N_{c1}/2)$ as in Figure 6.


\begin{figure}[htbp]
\setlength{\unitlength}{0.0125in}
\begin{picture}(332,110)(-80,-10)
\put(40.000,65.000){\arc{100.000}{0.6435}{2.4981}}
\put(120.000,5.000){\arc{100.000}{3.7851}{5.6397}}
\put(280.000,65.000){\arc{100.000}{0.6435}{2.4981}}
\put(200.000,35.000){\arc{80.000}{3.1416}{6.2832}}
\path(0,35)(320,35)
\put(0,15){$\mathbf{ c_2}$}
\put(80,15){$\mathbf{ c_1}$}
\put(160,15){$\mathbf{ 0}$}
\put(240,15){$\mathbf{ a_1}$}
\put(320,15){$\mathbf{ a_2}$}
\put(20,5){$\mathbf{ 6N_{f_2}}$}
\put(75,55){$\mathbf{ 3N_{f_1}+6N_{f_2}-N_{c_2}-2}$}
\put(155,75){$\mathbf{6N_{f_1} + 3 N_{f_2} -\frac{N_{c_1}}{2} +2 }$}
\put(260,5){$\mathbf{ 6N_{f_1}}$}
\end{picture}

\caption{}
\label{fig:test}
\end{figure}


In order to generalize to the case of arbitrary $k$, we take $(2k+1)$ 
copies of all
the singular points $a_{1}, 0$ and $ c_{1}$. When all these copies 
coincide, then 
we will have $(2k+1)N_{f2}$ D6 branes wrapped on $[c_{2}, a_1]$ and
$(2k+1)N_{f1}$ wrapped on $[c_1, a_{2}]$. 
The electric theory has the following
configuration of singular points $(a_{2}, a_{11} = \cdots = a_{1(2k+1)} 
= a_{1}, 
b_{1} = \cdots = b_{k} = 0 = b_{k+1} = \cdots =b_{2k+1}, 
c_{11} = \cdots = c_{1(2k+1)} = c_{1}, c_{2})$, 
with $2(2k+1)N_{f2}$ D6 branes wrapped 
on $[a_{2}, a_{1}]$, $((2k+1)N_{f2} + N_{c1}/2-2)$ D6 branes wrapped
on $[a_{1}, 0]$, $((2k+1)N_{f1} + N_{c2}+2)$ D6 branes wrapped on $[0, c_{1}]$
and $2(2k+1)N_{f1}$ D6 branes wrapped on $[c_{1}, c_{2}]$.   
The magnetic dual theory can be obtained by changing the position of $0$ 
and $c_{1}$ from
right to left with respect to $a_{1}$. 
Finally we  get the magnetic theory with the
gauge group $SO(2(2k+1)N_{f1} + 4(2k+1)N_{f2} - 2N_{c2})\times 
Sp(2(2k+1)N_{f1} + (2k+1) N_{f2} - N_{c1}/2)$ 
which again agrees with the one 
obtained in brane configuration for 
any value of $k$.

Let us discuss the duality for adjoint matter in the field theory setup. 
Besides the field $X$
that we had in the previous subsections we introduce adjoint fields for both
gauge groups of the product gauge group. We denote by $X_{1}(X_{2})$ the
antisymmetric(symmetric) adjoint
fields of $SO(N_{c1})(Sp(N_{c2}))$. The matter content of the electric theory
is given by (we will 
consider only the gauge groups and $U(1)_{R}$ charges for the matter
fields):
\vskip12pt

\begin{tabular}{|c|c|c|c|}
\hline
 &$SO(N_{c1})$&$Sp(N_{c2})$&$U(1)_{R}$\\
\hline
$Q^{(1)}$&${ \bf N_{c1}}$&{\bf 1}&
$1+\frac{1}{N_{f1}}(\frac{N_{c2}-N_{c1}}{k+1}+2)$
\\
\hline
$Q^{(2)}$&{\bf 1}&$ {\bf 2N_{c2}}$&$ 1+
\frac{1}{N_{f2}}(\frac{N_{c1}-4N_{c2}}
{2(k+1)} - 2)$\\
\hline
$A_{1}$&${\bf (N_{c1}-1)N_{c1}/2}$&{\bf 1}&$ \frac{k+2}{2(k+1)}$\\
\hline
$A_{2}$&{\bf 1}&${\bf 2N_{c2}(2N_{c2}+1)/2}$&$ \frac{k+2}{2(k+1)}$\\
\hline
 X&${\bf N_{c1}}$&${\bf 2N_{c2}} $&$ \frac{k}{2(k+1)}$\\
\hline
\end{tabular}

The superpotential for the theory is
$$
W = \mbox{Tr} A^{2(k+1)}_{1} + \mbox{Tr} A^{2(k+1)}_{2} + 
\mbox{Tr} A_{1} X^{2} - \mbox{Tr} A_{2} X^{2}.
$$
The chiral ring truncates because of this superpotential, in the
sense that the maximum power of $A_{i}$ that can appear is $k$. 
 The gauge invariant mesons in the theory are 
$$
Q^{(1)} A^{j}_{1}Q^{(1)}, Q^{(2)} A^{j}_{2} Q^{(2)},
Q^{(1)}A^{j}_{1} X Q^{(2)}, Q^{(1)} A^{j}_{2} X Q^{(2)}, 
Q^{(1)} X^{2} A^{j}_{1} Q^{(1)}, Q^{(2)} X^{2} A^{j}_{2} Q^{(2)}
$$
where $j=1, \cdots, k-1$.

The dual theory is described by an $SO(2(2k+1)N_{f1} + 
4(2k+1) N_{f2} - 2N_{c2})
\times Sp(2(2k+1)N_{f1} + (2k+1)N_{f2} - N_{c1}/2)$ gauge theory
with the following charged matter contents:

\begin{tabular}{|c|c|c|c|}
\hline
 &$SO(\widetilde{N}_{c1})$&$Sp(\widetilde{N}_{c2})$&$U(1)_{R}$\\
\hline
$q^{(1)}$&${\bf \widetilde{N}_{c1}}$&{\bf 1} &
$1+\frac{1}{N_{f1}}(\frac{\widetilde{N}_{c2}-\widetilde{N}_{c1}}{k+1}+2)$
\\
\hline
$q^{(2)}$&{\bf 1}&${\bf 2\widetilde{N}_{c2}} $&$ 
1+\frac{1}{N_{f2}}(\frac{\widetilde{N}_{c1}-
4\widetilde{N}_{c2}}
{2(k+1)} - 2)$\\
\hline
$\widetilde{A}_{1}$&${\bf (\widetilde{N}_{c1}-1)\widetilde{N}_{c1}/2} $
& {\bf 1}&
$\frac{k+2}{2(k+1)}$\\
\hline
$\widetilde{A}_{2}$&{\bf 1} &$ {\bf 2\widetilde{N}_{c2}(2\widetilde{N}_{c2}+
1)/2}$&$ \frac{k+2}{2(k+1)}$\\
\hline
 Y&${ \bf \widetilde{N}_{c1}}$&$ {\bf 2\widetilde{N}_{c2}}$&
$ \frac{k}{2(k+1)}$\\
\hline
\end{tabular}

where $\widetilde{N}_{c1}=2(2k+1)N_{f1} + 4(2k+1) N_{f2} - 2N_{c2}$ and
$\widetilde{N}_{c2}=2(2k+1)N_{f1} + (2k+1)N_{f2} - N_{c1}/2$.
In the dual theory, the images of the mesons coming from the electric 
field theory
are singlet fields and the dual superpotential is obtained by adding
coupling terms between singlets and dual mesons  to the superpotential of
the electric theory. 

\section{ $ SO(N_{c1}) \times Sp(N_{c2}) \times SO(N_{c3}) $ }

Since we have seen a brane configuration containing two sets of
D4 branes suspended between three NS5 branes, it is natural to ask what happens
for a brane configuration including more D4 branes suspended
between more than three NS5 branes. We expect
 to see {\it new} dualities in the 
context of  brane configuration  which
have not been studied before even at the field theory level.

\subsection{ The theory with simplest superpotential }

Let us study triple product gauge group for the electric theory 
for the simplest case where each of the four kinds of NS5 branes has 
its single copy. The gauge group is 
$SO(N_{c1})\times Sp(N_{c2}) \times SO(N_{c3})$
with $2N_{f1}(2N_{f2})[2N_{f3}]$ flavors in the vector ( 
fundamental)[vector] representation of the respective 
$SO(N_{c1})(Sp(N_{c2}))[SO(N_{c3})] $ gauge group.
We label them by letters A, B, C and D
from left to right on the compact $x^{6}$ direction.
The four NS5 branes are oriented at arbitrary angles in 
$(x^{4}, x^{5}, x^{8}, x^{9})$ directions. None of these NS5 branes are
parallel. 

There exist $N_{c1}/2$ D4 branes stretched between A 5 brane and 
B 5 brane which is connected to C 5 brane by $N_{c2}$
D4 branes. The C 5 brane is connected to D 5 brane by $N_{c3}/2$ 
D4 branes.
Between A(B)[C] 5 brane and B(C)[D] 5 brane 
we have $N_{f1}(N_{f2})[N_{f3}]$ D6 branes intersecting the 
$N_{c1}/2(N_{c2})[N_{c3}/2]$
D4 branes(plus their mirrors).

In order to find magnetic dual theory, it is easy to see 
from the similar previous arguments
that 
by moving all the 
$N_{f1}$  D6 branes to the left of all NS5 branes,
each D6 brane has three D4 branes on its left after   
transition and by moving all $N_{f3}$ D6 branes to the right,  
each D6 brane has three D4 branes on its left after transition.
First move B 5 brane  to the right of A 5 brane. 
Two D4 branes must
appear because we have $SO$ gauge group according to \cite{eva} or
two D4 branes must be stuck on the orientifold O4 plane in
the language of \cite{EGKRS}. We move C 5 brane to the right of 
A 5 brane. 
When C 5 brane passes A 5 brane, two D4 branes disappear between A 5 brane 
and C 5 brane because we have an $Sp$ gauge group,
so we have two D4 branes between C 5 brane and B 5 brane and no extra D4 branes
between A 5 brane and C 5 brane. Now move D 5 brane to the right of A 5 brane. 
When D 5 brane passes A 5 brane  there appear two new D4 branes between 
A 5 brane and D 5 brane. 
In this moment we have two supplementary D4 branes between A 5 brane and 
D 5 brane, no supplementary  D4 branes between D 5 brane and C 5 brane 
and two supplementary D4 branes between C 5 brane and B 5 brane.
Similarly we move  C 5 brane to the right of B 5 brane
there are two new D4 branes because of the $SO$ gauge group, but 
the previous supplementary two D4 branes will change their orientations,
they will cancel the new two D4 branes
which leads eventually to no supplementary or missing D4 branes between
B 5 brane and C 5 brane.
Now move D 5 brane to the right of B 5 brane, then we have minus two
D4 branes. After moving D 5 brane to the right of C 5 brane we finally get
plus(minus)[plus] two D4 branes between A(B)[C] 5 brane and B(C)[D] 5 brane.

The final configuration is the following, from left to right:
semiinfinite $N_{f2}$ D4 branes ending on A 5 brane and semiinfinite
$N_{f2}$ D4 branes ending on B 5 brane. There are $3N_{f1}$ between
$N_{f1}$ D6 branes and A 5 brane which is connected to B 5 brane.
There are $\widetilde{N}_{c3}/2$ D4 branes between A 5 brane and B 5 brane
which is connected to C 5 brane. We
have $\widetilde{N}_{c2}$  between B 5 brane and C 5 brane which is 
connected to D 5 brane. There are $\widetilde{N}_{c1}/2$ between C 5 brane
and D 5 brane which is connected to $N_{f3}$ D6 branes by $3N_{f3}$
D4 branes. There are $N_{f2}$ D6 branes to the right of $N_{f3}$ D6 branes.
Between C 5 brane and $N_{f2}$ D6 branes we have $2N_{f2}$ D4 branes and
between D 5 brane and $N_{f2}$ D6 branes there are $2N_{f2}$ D4 branes also.
Finally we have $2N_{f2}$ semiinfinite D4 branes ending on $N_{f2}$ D6
branes. Of course we have to
add their mirrors in the above configuration. 

The introduction of
semiinfinite D4 branes is made in the same spirit of \cite{bh} and
is necessary in order to be consistent with the result of 
geometric picture and of field theory which will be explained later.
As in \cite{ca3}, the appearance of semiinfinite D4 branes in the brane
configuration picture is the same as the fact that D6 branes must
wrap on cycles between {\it pairs of singular points}. We will see that
the results match if we introduce {\it exactly} this amount of
semiinfinite D4 branes. 

>From the linking number conservation of D 5 brane, we can read it before
the transition and after transition:
$-N_{c3}/2+N_{f3}/2+N_{f2}/2+N_{f1}/2=\widetilde{N}_{c1}/2-3N_{f3}-
2N_{f2}+N_{f3}/2+N_{f2}/2-2-N_{f1}/2$.
We obtain $\widetilde{N}_{c1}=2(N_{f1}+2N_{f2}+3N_{f3})-N_{c3}+4$.
Similarly the linking number of A 5 brane tells us that
$\widetilde{N}_{c3}=2(3N_{f1}+2N_{f2}+N_{f3})-N_{c1}+4$ from the condition,
$N_{c1}/2-N_{f1}/2-N_{f2}/2-N_{f3}/2=N_{f2}/2-N_{f1}/2+N_{f3}/2+N_{f2}-
\widetilde{N}_{c3}/2+3N_{f1}+2$. 
Finally, we get $\widetilde{N}_{c2}=2(N_{f1}+2N_{f2}+N_{f3})-N_{c2}-2$
from the relation of the linking number of B 5 brane, $
N_{c2}-N_{c1}/2-N_{f2}/2-N_{f3}/2+N_{f1}/2=
N_{f2}-\widetilde{N}_{c2}+\widetilde{N}_{c3}/2+N_{f2}/2+N_{f3}/2
-N_{f1}/2-4$.

Now we go to the field theory results.
The electric theory has the gauge 
group $SO(N_{c1})\times Sp(N_{c2})\times SO(N_{c3})$ with the following
matter content of fields :

\begin {tabular}{|c|c|c|c|c|c|}    
\hline
 &$SO(N_{c1})$&$Sp(N_{c2})$&$SO(N_{c3})$& $U(1)_{R}$\\
\hline
$Q^{(1)}$&${\bf N_{c1}}$&{\bf 1}&{\bf 1}&
$1+\frac{1}{2N_{f1}} (N_{c2}-N_{c1}+2)$\\
\hline
$Q^{(2)}$&{\bf 1}&${\bf 2N_{c2}}$&{\bf 1}&$1+\frac{1}{2N_{f2}}(\frac{N_{c1}+
N_{c3}}{2}-2-2N_{c2})$\\
\hline
$Q^{(3)}$&{\bf 1}&{\bf 1}&${\bf N_{c3}}$&$1+
\frac{1}{2N_{f3}} (2N_{c2}-N_{c3}+2) $\\
\hline
$X_{1}$&${\bf N_{c1}}$&${\bf 2N_{c2}}$&{\bf 1}&$\frac{1}{2}$\\
\hline
$X_{2}$&{\bf 1}&${\bf 2N_{c2}} $&${\bf N_{c3}}$&$\frac{1}{2}$\\
\hline
\end{tabular}   

The superpotential is given by
$$
W=\frac{1}{2} \mbox{Tr} X_{1}^{4(k+1)} + 
\mbox{Tr} X_{1}^{2(k+1)} X_{2}^{2(k+1)} + \frac{1}{2}
\mbox{Tr} X_{2}^{4(k+1)}.
$$
To simplify the charges of these fields under different symmetries and
to ease  
the writing of mesons of the theory, we took $k=0$ in the above
table.
The form of the superpotential will determine the restrictions on the powers
of $X_{1}, X_{2}$, i.e., $X_{1}^{5}=X_{2}^{5}=0$ which constrains  the
number of mesons which can be formed. 

The mesons in the theory are given by
\bea
& & M^{(1)}_{1}=Q^{(1)}Q^{(1)}, M^{(1)}_{2}=
Q^{(1)}X^{2}_{1}Q^{(1)}, M^{(1)}_{3}=Q^{(1)}X^{4}_{1}Q^{(1)}, \nonumber \\
& & M^{(2)}_{1}=Q^{(2)}Q^{(2)}, M^{(2)}_{2}=Q^{(2)}X^{2}_{1}Q^{(2)},
M^{(2)}_{3}=Q^{(2)}X^{4}_{1}Q^{(2)}, M^{(2)}_{4}=Q^{(2)}X^{2}_{2}Q^{(2)}, 
\nonumber \\
& & M^{(3)}_{0}=Q^{(3)} Q^{(3)}, 
 M^{(3)}_{1}=Q^{(3)}X^{2}_{2} Q^{(3)}, 
 M^{(3)}_{2}=Q^{(3)}X^{4}_{2} Q^{(3)}. \nonumber
\eea
The dual group is $SO(\widetilde{N}_{c1})\times Sp(\widetilde{N}_{c2})
\times SO(\widetilde{N}_{c3})$ where $\widetilde{N}_{c1} = 2(
3N_{f3} + 2N_{f2}+N_{f1}) - N_{c3} + 4, \widetilde{N}_{c2} = 
4N_{f2} + 2N_{f1} + 2N_{f3}
- N_{c2} - 2, \widetilde{N}_{c3} = 2(3N_{f1} + 2N_{f2} + N_{f3}) - 
N_{c1}+ 4$.  
The content of the fields in the dual theory with their charges
is as follows(the flavor group is 
the same in the dual theory as in the electric
theory which is a common feature of $N=1$ dualities):

\begin{tabular}[b]{|c|c|c|c|c|c|}
\hline
 &$SO(\widetilde{N}_{c1})$&$Sp(\widetilde{N}_{c2})$&$SO(
\widetilde{N}_{c3})$&$U(1)_{R}$\\
\hline
$q^{(1)}$&${\bf \widetilde{N}_{c1}}$&{\bf 1}&{\bf 1}&$1+\frac{1}{2N_{f1}}
(\widetilde{N}_{c2}-\widetilde{N}_{c1}+2)$\\
\hline
$q^{(2)}$&{\bf 1}&${ \bf 2\widetilde{N}_{c2}}$&{\bf 1}&$1+\frac{1}{2N_{f2}}
(\frac{\widetilde{N}_{c1}+\widetilde{N}_{c3}}{2}-2-2\widetilde{N}_{c2})$\\
\hline
$q^{(3)}$&{\bf 1}&{\bf 1}&${\bf \widetilde{N}_{c3}}$&$1+\frac{1}{2N_{f3}}
(2\widetilde{N}_{c2}-\widetilde{N}_{c3}+2)$\\
\hline
$Y_{1}$&${ \bf \widetilde{N}_{c1}}$&${\bf 2\widetilde{N}_{c2}}$&
{\bf 1}&$\frac{1}{2}$\\
\hline
$Y_{2}$&{\bf 1}&${\bf 2\widetilde{N}_{c2}}$&${\bf 
\widetilde{N}_{c3}}$&$\frac{1}{2}$\\
\hline
\end{tabular}

The fundamental quarks have reversed their global
symmetries in the dual theory. There is a one-to-one map from the mesons of
the original theory to the singlets of the dual theory. The singlets of
the dual theory will enter in the superpotential which appear in the dual.
Other checks of this duality can be performed. The t' Hooft anomaly
matching conditions and 
the flow to other dualities by giving expectation values to
different flavors are satisfied.
The generalization to other values for $k$ goes in the same way, the difference
being the number of mesons which increases. 

\subsection{ The addition of adjoint matter}

Now we introduce adjoint matter.
In the brane configuration picture, 
 this corresponds 
to $(2k+1)$ NS5 branes with the same orientation as A connected by $N_{c1}$
D4 branes with $(2k+1)$ NS5 branes with the same orientation as B. 
These are
connected by $N_{c2}$ D4 branes with $(2k+1)$ NS5 branes 
with the same orientation
as C and the last ones are connected by $N_{c3}$ D4 branes 
with $(2k+1)$ NS5 branes
with the same orientation as D. 

>From the linking number conservation of D 5 brane, we can read it before
the transition and after transition:
$-N_{c3}/2+(2k+1)N_{f3}/2+(2k+1)N_{f2}/2+(2k+1)N_{f1}/2=
\widetilde{N}_{c1}/2-3(2k+1)N_{f3}-
2(2k+1)N_{f2}+(2k+1)N_{f3}/2+(2k+1)N_{f2}/2-2-(2k+1)N_{f1}/2$.
We obtain $\widetilde{N}_{c1}=2(2k+1)(N_{f1}+2N_{f2}+3N_{f3})-N_{c3}+4$.
Similarly the linking number of A 5 brane tells us that
$\widetilde{N}_{c3}=2(2k+1)(3N_{f1}+2N_{f2}+N_{f3})-N_{c1}+4$ 
from the condition
$N_{c1}/2-(2k+1)N_{f1}/2-(2k+1)N_{f2}/2-(2k+1)N_{f3}/2=
(2k+1)N_{f2}/2-(2k+1)N_{f1}/2+(2k+1)N_{f3}/2+(2k+1)N_{f2}-
\widetilde{N}_{c3}/2+3(2k+1)N_{f1}+2$. 
Finally we get $\widetilde{N}_{c2}=2(2k+1)(N_{f1}+2N_{f2}+N_{f3})-N_{c2}-2$
from the relation of the linking number of B 5 brane $
N_{c2}-N_{c1}/2-(2k+1)N_{f2}/2-(2k+1)N_{f3}/2+(2k+1)N_{f1}/2=
(2k+1)N_{f2}-\widetilde{N}_{c2}+\widetilde{N}_{c3}/2+(2k+1)N_{f2}/2+
(2k+1)N_{f3}/2
-(2k+1)N_{f1}/2-4$.

It is possible to obtain the duality for the 3-product with 
adjoint matter at the field theory level.
For the same product gauge group, we introduce the fields $A_{1}(A_{2})[A_{3}]$
 in the antisymmetric(symmetric)[antisymmetric] 
adjoint representation of $SO(N_{c1})(Sp(N_{c2}))[SO(N_{c3})]$ besides
the fields, $X_1, X_2$ 
that were in the previous subsection. Again we truncate the chiral 
ring by adding the superpotential:
$$
W = \frac{1}{k + 1} \mbox{Tr} A^{k+1}_{i} + \mbox{Tr} A_{1} X^{2}_{1}
+ \mbox{Tr} A_{2} X^{2}_{2} + \mbox{Tr} A_{2} X^{2}_{1} - 
\mbox{Tr} A_{3} X^{2}_{2}, \;\;\; i=1,2,3 
$$
The conditions for supersymmetric vacuum reduce to the number of mesons.
The dual group is $SO(2(2k+1)N_{f1} + 4(2k+1) N_{f2} + 6(2k+1) N_{f3} -
N_{c3} + 4)\times Sp (2(2k+1)N_{f1} + 4(2k+1) N_{f2} + 2(2k+1) N_{f3}
- N_{c2} - 2) \times SO(6(2k+1)N_{f1} + 4(2k+1)N_{f2} + 2(2k+1) N_{f3}
- N_{c1} + 4)$. Again the flavors transform under different groups
compared with the electric theory. The mesons of the electric theory
become now singlets and will enter in the superpotential for the
magnetic superpotential together with the dual mesons.

\section {  $ SO(N_{c1}) \times Sp(N_{c2}) \times SO(N_{c3}) \times 
Sp(N_{c4}) $ }

We want to see {\it new} dualities in the 
context of brane configurations which
have not been studied before.
Let us study 4-tuple product gauge group for the electric theory 
for the simplest case in this section where five NS5 branes have its single
copy. 

\subsection{The theory with simplest superpotential}

The gauge group is 
$SO(N_{c1})\times Sp(N_{c2}) \times SO(N_{c3}) \times Sp(N_{c4})$
with $2N_{f1}(2N_{f2})[2N_{f3}]\{2N_{f4}\}$ flavors in the 
vector(fundamental)[vector]\{fundamental\} representation of respective 
$SO(N_{c1})(Sp(N_{c2}))[SO(N_{c3})]\{Sp(N_{c4})\} $ gauge group.
We label them by letters A, B, C, D and E 
from left to right on the compact $x^{6}$ direction.
The five NS5 branes are oriented at arbitrary angles in 
$(x^{4}, x^{5}, x^{8}, x^{9})$ directions. None of these NS5 branes are
parallel. 

There exist $N_{c1}/2$ D4 branes stretched between A 5 brane and 
B 5 brane which is connected to C 5 brane by $N_{c2}$
D4 branes. The C 5 brane is connected to D 5 brane by $N_{c3}/2$ 
D4 branes. The D 5 brane is connected to E 5 brane by $N_{c4}$
D4 branes.
Between A(B)[C]$\{$D$\}$ 5 brane and B(C)[D]$\{$E$\}$ 5 brane 
we have $N_{f1}(N_{f2})[N_{f3}]\{N_{c4}\}$ 
D6 branes intersecting the 
$N_{c1}/2(N_{c2})[N_{c3}/2]\{N_{f4}\}$
D4 branes(plus their mirrors).

In order to go to magnetic dual theory,  
by moving all the 
$N_{f1}$  D6 branes to the left of all NS5 branes,
each D6 brane has four D4 branes on its left after   
transition and by moving all $N_{f4}$ D6 branes to the right,  
each D6 brane has four D4 branes on its right.
First move B 5 brane  to the right of A 5 brane. 
Two D4 branes must
appear because we have $SO$ gauge group. We move C 5 brane to the right of 
A 5 brane. 
When C 5 brane passes A 5 brane, two D4 branes disappear between A 5 brane 
and C 5 brane because we have an 
$Sp$ gauge group,
so we have two D4 branes between C 5 brane and B 5 brane 
and no extra D4 branes
between A 5 brane and C 5 brane. Now move D 5 brane to the right of A 5 brane. 
When D 5 brane passes A 5 brane  there are two new D4 branes between them 
because of the $SO$ gauge group.
Similarly we move  C 5 brane to the right of B 5 brane
there are two new D4 branes but the previous two D4 branes are changing the
orientation so they annihilate the two new D4 branes which leads eventually 
to no D4 branes created or annihilated. 
The gauge group in the magnetic theory is
then given by $SO(\widetilde{N}_{c1})\times 
Sp(\widetilde{N}_{c2})\times SO(\widetilde{N}_{c3})
\times Sp(\widetilde{N}_{c4})$ with inverted order of the corresponding
flavors.
As we did before, the values for $\widetilde{N}_{i}(i=1, 2, 3, 4)$ 
are calculated by linking
number conservations and we obtain the following values:  
$\widetilde{N}_{c1}=2( N_{f1}+2N_{f2}+3N_{f3}+4N_{f4})-2N_{c4},
\widetilde{N}_{c2}=2N_{f1}+4N_{f2}+6N_{f3}+3N_{f4}-N_{c3}/2,
\widetilde{N}_{c3}=2( 3N_{f1}+6N_{f2}+4N_{f3}+2N_{f4})-2N_{c2},
\widetilde{N}_{c4}= 4N_{f1}+3N_{f2}+2N_{f3}+N_{f4}-N_{c1}/2$.

\section{  Generalization to higher product gauge groups}

We have seen that a number of $N=1$ supersymmetric field theory dualities
were obtained in terms of brane configuration and
geometric realization of wrapping D6 branes
around 3-cycles of CY threefold in type IIA string theory.
Our construction 
also gives rise to extra D6 branes
wrapping around the cycles.
It would be interesting to study this transition further.  
The above construction can be generalized to any product of gauge groups, 
but we have
to put them in alternating order, i.e., $\cdots 
SO(N_{c1})\times Sp(N_{c2})\times 
SO(N_{c3})\times Sp(N_{c4})\times \cdots $(by changing the overall sign of
$\Omega^{2}$ we can start with a $Sp$ gauge group from left to right).

In order to obtain a generalization to any value of $n$, one has to take 
$(2n-1)$ singular points and to move them from left to right with respect to
a reference point. We always have to be able to extend the flavor cycles to 
infinity. So we always push $a_{2}$ to the far right and $c_{2}, d_{2},
\cdots$
to the far left. In order to obtain the result of field theory, we 
have to twist the fibration in order to move 
 the D 6-branes which do not contribute to the
gauge group to the desired direction.
This is equivalent to   introducing the semi-infinite
D 4-branes in brane configuration method. 
Our result again agrees with the one
of \cite{bh}.

For a product of more than two gauge groups, there are two cases:

1) when there is an even number of gauge groups in the product, i.e.,
$\prod_{i=1} SO(N_{ci}) \times Sp(N_{c(i+1)})$ where $i=2j-1$ and $j=1, 2,
\cdots$ 
the effects of $SO$
and $Sp$ projections will cancel each other so in the overall
picture of the dual, no D4 branes appear or disappear. 
The result is similar to the
one obtained by \cite{bh} with the modifications that are to be done when
one considers non-orientable string theory. Taking their results, we 
simply modify
$\widetilde{N}_{c}$ to $2\widetilde{N}_{c}$ and $N_{c}$ to $2N_{c}$ 
anytime we talk
about the $Sp$ gauge groups, obtaining the dual theory  
for the alternating product
of $SO$ and $Sp$ gauge groups. The argument that we use for the product of
two gauge groups applies also here. So one has to be careful when changing
the positions of two NS5 branes connected by supplementary D4 branes or having
a deficit of D4 branes between them. 

2) when there is an odd number of gauge groups in the product, i.e.,
$(\prod_{i=1} SO(N_{ci}) \times Sp(N_{c(i+1)})) \times SO(N_{c})$
where $i=2j-1$ and  $j=1, 2, \cdots,$   
we need to 
create or to annihilate D4 branes in the overall picture(or 
to put D4 branes or anti D4 branes on
the top of the orientifold in order to make a smooth transition).

\section{Conclusions}

Originally, the dualities for N=1 supersymmetric theories were 
considered between theories which had the same quantum chiral
ring and moduli space of vacua as a function of possible deformations
determined by the expectation values of the scalar fields of
both theories. It is very difficult to find the dual theory 
because nobody knows a specific recipe for obtaining it. The models
were considered separately and only some examples which enter in
a general scheme have been obtained. By embedding the problem in string theory
duality is translated in brane mechanics. The classical moduli space
for both theories are embedded in a single moduli space of string vacua
and this makes the relation between the two moduli space manifest.
The quarks, antiquarks and mesons have a natural interpretation as 
different strings forced to end on different D branes and as oscillations
of the specific D4 branes. The superpotential for the specific models
considered so far appears from their $N=2$ origin after rotating
NS5 branes. 

In the geometrical picture, the relation between the moduli spaces of both
theories becomes again manifest. In this approach, we just go between 
different points of the moduli space of Calabi-Yau manifolds. 
By wrapping D6 branes on the vanishing three cycles on the doubly elliptic
fibered Calabi-Yau manifold, we realize $N=1$ dualities in four dimensions
for various gauge groups. Via T-duality, we connect these geometric
configurations to the brane configurations.
 In brane configuration picture, semi-infinite branes are to be
introduced and we always have to check that the result is identical with the
one obtained in geometric picture and by field theory methods.

Many results have been obtained before for different gauge groups and
several matter representations.
In the present work we generalized the work of \cite{Tatar, ca3}
to the case of product gauge groups between $SO$ and $Sp$ gauge groups.
The main ideas, in the geometric picture, were the appearance of 
supplementary D6 branes when
the positions of two singular points are inverted and the fact that
in magnetic dual 
theory the flavor groups are given by D6 branes wrapped on cycles
between {\it pairs} of singular points. In brane configuration picture the
last condition was translated in the appearance of semi-infinite
D4 branes as in \cite{bh}.

All these $N=1$ configurations are obtained from $N=2$ configurations
by rotating the NS5 branes. Considering the fact that in $N=2$
theories we cannot have $SO(N_{c})$ gauge group with $SO(N_{f})$
flavor group, this implies that, by gauging the flavor group, we cannot
have  $SO(N_{c})\times SO(N_{f})$ gauge group. 
Using the arguments of \cite{eva, EGKRS}, this can be understood
by using the fact that the charge of the O4 plane is connected with the
sign of $\Omega^{2}$ so for each NS5 brane, we must have $SO$ projection
on one side and $Sp$ projection on the other side.
Therefore this brane configuration seems not to be suitable for
describing the gauge theory with $SO(N_{c1})\times SO(N_{c2})$ gauge
group or $Sp(N_{c1})\times Sp(N_{c2})$ gauge group.
As we know at the field theory level these dual relations have
been found already in the paper of \cite{ils}. It would be interesting
to study how these dualities arise from  brane configuration
or geometric configuration.  Another very interesting case 
is the gauge theory with $SO(N_{c1})\times SU(N_{c2})$ where 
the orientifold should act only on one part of the configuration.
It is again clear that the presently known brane configuration
cannot describe this model. Matter in the spinor representation
would allow us to obtain other very interesting model,
especially the duality between chiral theories with $SO$ group and
non-chiral theories with $SU$  group. All this possible ideas
are urging us to find a new configuration preserving $N=1$ in
$d=4$ dimensions. 

In the beautiful lectures of Townsend 
given at Cargese School 1997, 
there appear to be few configurations which
preserves 1/8 of supersymmetries.
In order to obtain an effective four-dimensional theories, some
D branes should have compact directions such that the world seen
by an observer who lives on them is a four-dimensional
field theory. In all the cases considered so far, D4 branes with
one compact directions were considered.
It seems that some possibilities
for other interesting configurations would be to consider D4 branes
which are compact in more that one direction, like a D5 brane
ending on D7 brane and with two compact directions. It should be
extremely interesting to obtain new brane configurations which
preserve 1/8 of supersymmetry and simultaneously 
which cannot be obtain from the
configurations which preserve 1/4 of supersymmetry.
 
Therefore we expect many new developments in this line of research
which will give us many new insights into field theory dualities
and also string theories with a better understanding of brane
configurations. Developments in both theories have a common goal of
obtaining information about the strong coupled sectors of both 
theories.

{\bf Acknowledgments}

We thank Jaemo Park for discussions 
who participated in early stage of this work.
C.A. thanks for the hospitality of APCTP where this work has been done.
R.T. would like to thank to the organizers of Cargese Summer School for
the extremely stimulating atmosphere created during the entire school
and John Brodie, Per Kraus, Adam Schwimmer and Zheng Yin for very useful 
discussions. K. Oh would like to thank C. Vafa for e-mail correspondences.

\section{Appendix}

We give here a short explanation for the new phenomenon occurring in 
the transition from the electric to magnetic theory in the OV approach.
Let us discuss for the orientable case, i.e., with $SU$ groups.
Consider a NS5 brane denoted by A in (12345) directions, another NS5 brane
denoted by B in (12389) directions and $N_{f}$ D4 branes between them in (1236)
directions. We do not have any D6 branes. We want to move the B 5 
brane from the
left to the right of the A 5 brane. 
Because we do not have any D6 brane, we cannot
make the transition to the Higgs phase. So, in order to preserve supersymmetry,
we are not allowed to move the B 5 brane in the $x^{7}$ direction. In order
to pass it to the left of the A 5 brane we have to pass the NS5 branes one over
the other. But the D4 branes are bound to the 5 brane B and they will pass
through the stationary NS5 brane. In order to see better the phenomenon,
we consider that the NS5 branes are very closed to each other, so that the
extension of the D4 branes in the $x^{6}$ direction is negligible. 
Then we make a succession of dualities :

1) T-duality on (123) direction; the D4 brane becomes D1 brane(6)  but
$x^{6}$ is negligible so
it can be consider to be just a point stuck on the NS branes. 

2) S - duality so we have D5 brane(12345) and D5 brane(12389).

3) T-duality on $x^7$ direction so we obtain D6 brane(123457) and 
D6 brane(123789).

So we have one D6 brane (123457) passing through a D6 brane (123789).
If the (12367) space were a 5 torus, then our problem would reduce to the
one of \cite{BDG}. In that case, formula (6) of \cite{BDG} tells us that
the net charge inflow is $k$ where $k$ is the instanton number. Here
the instanton number is given by the charge of the D4 branes which became
points after the first three T-dualities. So the net charge inflow is
just $N_{f}$. This tells us that there is a fundamental string in the
$x^{6}$ direction which is created after the NS5 branes are passing through
each other. If we go backwards with the three steps of dualities, we obtain
from the F1 string(6) one D6 brane(1236). So, if the A and B NS5 branes are
passing each other, there are $N_{f}$ D4 branes that are created between the
two NS5 branes. The argument is not changed if we take the volume of the
5 torus to be very big and we go to the noncompact limit.
This is the phenomenon that occurs in the OV approach.

In the non-orientable case everything goes the same, but we have to count 
always the number of D4 branes as the sum of physical D4 branes
plus their mirrors.


\begin{thebibliography}{99}
\bibitem{KV1} S. Katz and C. Vafa, 
Geometric Engineering of N=1 Quantum Field Theories,
hep-th/9611090.
\bibitem{BJPSV} M. Bershadsky, A. Johansen, T. Pantev, V. Sadov and 
C. Vafa,
F-theory, Geometric Engineering and N=1 Dualities,
hep-th/9612052.
\bibitem{HW} A. Hanany and E. Witten, 
Type IIB Superstrings, BPS Monopoles, And Three-Dimensional Gauge Dynamics,
hep-th/9611230.
\bibitem{BDG} C.P. Bachas, M.R. Douglas and M.B. Green, 
Anomalous Creation of Branes,
hep-th/9705074.
\bibitem{DFK} U. Danielsson, G. Ferretti and I.R. Klebanov ,
Creation of Fundamental Strings by Crossing D-branes,
hep-th/9705084.
\bibitem{BGL} O. Bergman, M.R. Gaberdiel and G. Lifschytz,
Branes, Orientifolds and the Creation of Elementary Strings,
hep-th/9705130.
\bibitem{deAl} S.P. de Alwis, 
A note on brane creation,
hep-th/9706142.
\bibitem{kuta} S. Elitzur, A. Giveon and D. Kutasov, 
Branes and N=1 Duality in String Theory,
hep-th/9702014.
\bibitem{Seiberg} N. Seiberg, 
Electric-Magnetic Duality in Supersymmetric Non-Abelian Gauge Theories,
hep-th/9411149.
\bibitem{eva} N. Evans, C.V. Johnson and A.D. Shapere,  
Orientifolds, Branes, and Duality of 4D Gauge Theories,
hep-th/9703210.
\bibitem{bh} J.H. Brodie and A. Hanany, 
Type IIA Superstrings, Chiral Symmetry, and N=1 4D Gauge Theory Dualities,
hep-th/9704043.
\bibitem{BSTY} A. Brandhuber, J. Sonnenschein, S. Theisen and
S. Yankielowicz, 
Brane Configurations and 4D Field Theory Dualities,
hep-th/9704044.
\bibitem{EGKRS} S. Elitzur, A. Giveon, D. Kutasov, E. Rabinovici and
A. Schwimmer, 
Brane Dynamics and N=1 Supersymmetric Gauge Theory
hep-th/9704104.
\bibitem{Tatar} R. Tatar, 
Dualities in 4D Theories with Product Gauge Groups from Brane Configurations,
hep-th/9704198.
\bibitem{bar} J.L.F. Barbon,
Rotated Branes and N=1 Duality,
hep-th/9703051.
\bibitem{John} C.V. Johnson, 
On the Orientifolding of Type II NS-Fivebranes,
hep-th/9705148.
\bibitem{John1} C.V. Johnson, 
>From M-theory to F-theory, with Branes,
hep-th/9706155.
\bibitem{HZ} A. Hanany and A. Zaffaroni, 
Chiral Symmetry from Type IIA Branes,
hep-th/9706047.
\bibitem{IS} K. Intriligator and N. Seiberg, 
Duality, Monopoles, Dyons, Confinement and Oblique Confinement in 
Supersymmetric $SO(N_c)$ Gauge Theories,
hep-th/9503179.
\bibitem{IP} K. Intriligator and P. Pouliot, 
Exact Superpotentials, Quantum Vacua and Duality in 
Supersymmetric $SP(N_c)$ Gauge Theories,
hep-th/9505006.
\bibitem{ov} H. Ooguri and C. Vafa, 
Geometry of N=1 Dualities in Four Dimensions,
hep-th/9702180.
\bibitem{BSV} M. Bershadsky, V. Sadov and C. Vafa, 
D-Strings on D-Manifolds,
hep-th/9510225.
\bibitem{CDFV} A. Ceresole, R. D'Auria, S. Ferrara and A. Van Proeyen,
Duality Transformations in Supersymmetric Yang-Mills 
Theories coupled to Supergravity,
hep-th/9502072.
\bibitem{ca1} C. Ahn and K. Oh, 
Geometry, D-Branes and N=1 Duality in Four Dimensions I,
hep-th/9704061.
\bibitem{ca2} C. Ahn, 
Geometry, D-Branes and N=1 Duality in Four Dimensions II,
hep-th/9705004.
\bibitem{ca3} C. Ahn and R. Tatar, 
Geometry, D-branes and N=1 Duality in Four Dimensions with Product Gauge 
Group,
hep-th/9705106.
\bibitem{Kutasov} D. Kutasov, 
A Comment on Duality in N=1 Supersymmetric Non -- Abelian Gauge Theories,
hep-th/9503086.
\bibitem{ks} D. Kutasov and A. Schwimmer,
On Duality in Supersymmetric Yang-Mills Theory,
hep-th/9505004.
\bibitem{intril} K. Intriligator, 
New RG Fixed Points and Duality in Supersymmetric $SP(N_c)$ and $SO(N_c)$ 
Gauge Theories,
hep-th/9505051.
\bibitem{ls} R.G. Leigh and M.J. Strassler, 
Duality of Sp(2N) and SO(N) Supersymmetric Gauge Theories with Adjoint Matter,
hep-th/9505088. 
\bibitem{BS} J.H. Brodie and M.J. Strassler, 
Patterns of Duality in N=1 SUSY Gauge Theories,
hep-th/9611197.
\bibitem{Brodie} J. Brodie, 
Duality in Supersymmetric SU($N_c$) Gauge Theory with 
Two Adjoint Chiral Superfields,
hep-th/9605232.
\bibitem{Witten} E. Witten, 
Solutions Of Four-Dimensional Field Theories Via M Theory,
hep-th/9703166.
\bibitem{L3} K. Landsteiner, E. Lopez and D.A. Lowe, 
N=2 Supersymmetric Gauge Theories, Branes and Orientifolds,
hep-th/9705199.
\bibitem{BSTY1} A. Brandhuber, J. Sonnenschein, S. Theisen and 
S. Yankielowicz, 
M Theory And Seiberg-Witten Curves: Orthogonal and Symplectic Groups,
hep-th/9705232.
\bibitem{M3} A. Marshakov, M. Martellini and A. Morozov, 
Insights and Puzzles from Branes: 4d SUSY Yang-Mills from 6d Models,
hep-th/9706050.
\bibitem{FS} A. Fayyazuddin and M. Spalinski, 
The Seiberg-Witten Differential From M-Theory,
hep-th/9706087.
\bibitem{AH} O. Aharony and A. Hanany, 
Branes, Superpotentials and Superconformal Fixed Points,
hep-th/9704170.
\bibitem{bk} I. Brunner and A. Karch, 
Branes and Six Dimensional Fixed Points,
hep-th/9705022.
\bibitem{kol} B. Kol, 
5d Field Theories and M Theory,
hep-th/9705031.
\bibitem{HOO} K. Hori, H. Ooguri and Y. Oz, 
Strong Coupling Dynamics of Four-Dimensional N=1 Gauge Theories from M 
Theory Fivebrane,
hep-th/9706082.
\bibitem{bra} A. Brandhuber, N. Itzhaki, V. Kaplunovsky, J. Sonnenschein
and S. Yankielowicz, 
Comments on the M Theory Approach to N=1 SQCD and Brane Dynamics,
hep-th/9706127.
\bibitem{witt1} E. Witten, 
Branes And The Dynamics Of QCD,
hep-th/9706109.
\bibitem{hov} K. Hori, H. Ooguri and C. Vafa , 
Non-Abelian Conifold Transitions and N=4 Dualities in Three Dimensions,
hep-th/9705220.
\bibitem{cg} C. Gomez, 
On the geometry of soft breaking terms and N=1 superpotentials,
hep-th/9706131.
\bibitem{ils} K. Intriligator, R.G. Leigh and M.J. Strassler, 
New Examples of Duality in Chiral and Non-Chiral Supersymmetric 
Gauge Theories,
hep-th/9506148.
\bibitem{polchin} J. Polchinski, 
TASI Lectures on D-Branes,
hep-th/9611050. 
\end{thebibliography}
\end{document}